# Scanning tunneling spectroscopy of superconductivity on surfaces of LiTi$_2$O$_4$(111) thin films

Yoshinori Okada[1], Yasunobu Ando[2], Ryota Shimizu[1], Emi Minamitani[2], Susumu Shiraki[1], Satoshi Watanabe[2] & Taro Hitosugi[1,3]

[1]Advanced Institute for Materials Research, Tohoku University, Sendai 980-8577, Japan

[2]Department of Materials Engineering, The University of Tokyo, Tokyo 113-8656, Japan

[3]School of Materials and Chemical Technology, Tokyo Institute of Technology, Tokyo 152-8552, Japan

**Abstract**

Unique superconductivity at surfaces/interfaces, as exemplified by LaAlO$_3$/SrTiO$_3$ interfaces, and the high transition temperature in ultrathin FeSe films, have triggered intense debates on how superconductivity is affected in atomic and electronic reconstructions. The surface of superconducting cubic spinel oxide LiTi$_2$O$_4$ is another interesting system because its inherent surface electronic and atomic reconstructions add complexity to superconducting properties. Investigations of such surfaces are hampered by the lack of single crystals or high-quality thin films. Here, using low-temperature scanning tunneling microscopy, we report an unexpected small superconducting energy gap and a long coherence length on the surface of LiTi$_2$O$_4$ (111) epitaxial thin films. Furthermore, we find that a pseudogap opening at the Fermi energy modifies the surface superconductivity. Our results open an avenue, exploring anomalous superconductivity on the surface of cubic transition-metal oxides where the electronic states are spontaneously modulated with involving rich many-body interactions.

# Introduction

Recent advances in the atomic-scale synthesis and characterization of two-dimensional superconductors have kindled significant interest in their exotic electronic, orbital, and magnetic structures[1–6]. In addition to the ultrathin superconducting films on substrates[7–20], surfaces of cubic-structure superconducting transition-metal oxides provide another interesting platform, because the spontaneous electronic and atomic reconstructions on surfaces are expected to modify superconductivity. A particularly interesting system is the spinel oxide ($AB_2O_4$) superconductor. In this system, the cubic pyrochlore sub-lattice of $B$-atoms provides large degeneracy (frustration) of charge-, spin-, and orbital-states in bulk[21,22], and a prominent degeneracy lifting at the surface is expected to lead rich electronic states on the surface. None of the previous studies, however, has revealed the electronic signature of modulated superconductivity on their surfaces.

Lithium titanate, $LiTi_2O_4$, is the only oxide superconductor with spinel structure[23–26], and exhibits the highest superconducting transition temperature ($T_c$) with ~13.7 K of any spinel superconductors[27-32]. It is known as a $3d^{0.5}$ metallic system (one half of an electron per Ti atom resides in the 3d states), and medium-coupling Bardeen–Cooper–Schrieffer (BCS) superconductivity with *s*-wave pairing symmetry has been proposed[33–36]. Recent transport data with epitaxial films showed an angle-dependent anomalous magnetoresistance, which is possibly related to a spin and orbital fluctuation effect[22]. Based on this data, the similarity between $LiTi_2O_4$ and high-$T_c$ cuprates has been discussed[22]. Although superconducting properties of bulk $LiTi_2O_4$ have been investigated extensively, studies of the superconductivity at $LiTi_2O_4$ surfaces have been hindered by the lack of single crystals and high-quality thin films. Recently, high-quality epitaxial $LiTi_2O_4$ thin films were successfully grown using pulsed laser deposition (PLD)[37–39]; however, investigations of the superconductivity on $LiTi_2O_4$ surfaces remain unexplored. Accordingly, we have developed a scanning tunneling microscope combined with a PLD system[40] to examine details of the $LiTi_2O_4$ surface in situ without exposing the surface to air.

Here, we report modified superconductivity at the surfaces of $LiTi_2O_4$(111) epitaxial thin films, using scanning tunneling microscopy/spectroscopy (STM/STS) and first-principles density-functional theory (DFT) calculations. From the atomic-scale observations of superconductivity, we present measurements of the unexpected small superconducting energy gap $\Delta$ and a long coherence length $\xi$ values. Furthermore, we found that a superconducting gap exists on large energy scale pseudogap states. These results provide the spectroscopic evidence of spontaneous superconducting modification on the surface of cubic transition metal oxides, paving an interesting path to explore exotic superconductivity with involving rich interactions.

## Results

**Preparation of higher $T_c$ sample with flat surface.** We found that low-temperature film deposition using PLD followed by post-deposition annealing increases $T_c$ up to 13 K, together with improved surface flatness. We first compare the STM images of the films grown at substrate temperature $T_s$ of 600 ºC (Fig. 1a) and 400 ºC (Fig. 1b). The former temperature is the typical growth temperature reported in the previous studies [22, 37, 38, 39]. While many triangular-shaped islands are observed on both surfaces, the height of the islands are much smaller in the latter film, indicating that the surface roughness is strongly dependent on the growth temperatures. As the $T_s$ is reduced from 600 to 400ºC, the root mean square (RMS) value of surface roughness, $R_{RMS}$, improved from 0.86 (Fig. 1a) to 0.40 nm (Fig. 1b). To further improve the surface roughness, we investigated the post-deposition annealing effect on the film grown at $T_s = 400°C$; thin films were annealed at 600°C in UHV for an hour. Expectedly, the film became flatter to $R_{RMS} = 0.28$ nm (Fig. 1c). Furthermore, this $R_{RMS}$ value was improved to 0.21 nm for the film deposited at $T_s = 300°C$ followed by the above-mentioned annealing (Fig. 1d and Supplementary Fig. 1).

Importantly, this annealing process increases $T_c$ values (Fig. 1f), in addition to the flattening of the surfaces. While the $T_c$ values of the films grown at $T_s = 500°C$ and 600°C show 12.0 K and 12.5 K, respectively, the $T_c$ increases up to 13.0 K for the films after annealing (red curves in Fig. 1f). We stress that the $T_c$ value of ~13 K is one of the highest values among reported values for LiTi$_2$O$_4$ epitaxial thin films [22, 37, 38, 39]. We confirmed from XRD that the films exhibit rocking-curve full-width at half-maximum of 0.36° for 444 peak, and that the lattice is fully relaxed from SrTiO$_3$(111) substrate (Supplementary Fig. 2). The above results show that the quality of the sample is quite high, and the concentration of defects in the crystal is low. In the following investigations, we focus on the films deposited at $T_s = 300°C$ followed by the post-deposition annealing in vacuum at 600°C.

**Atomic-scale order on LiTi$_2$O$_4$ (111) film surface.** We observed, on the atomic level, a well-ordered triangular lattices and defects (Fig. 2). Fig. 2a shows the wide-area STM image of the LiTi$_2$O$_4$ surface, where flat terraces and dark spots with peculiar defect shapes were observed. In the close-up image (Fig. 2c), strong topographic corrugations (shown in blue) are observed together with weak topographic corrugations (shown in red), displaying 3-fold symmetry. This is consistent with the inherent 3-fold rotational symmetry along the [111] axis in a spinel system. The unit cell size was found to be ~0.6 nm (Fig. 2c), which was independent of the location and sample bias voltages $V_s$. This size well matches that of the unit cell of $LiTi_2O_4$ bulk (111) plane. We concluded that the protrusions of the triangular lattice originate from Ti atoms on the surface, as discussed later. Also, nature of the defects (dark spots) is addressed later.

**Superconductivity in tunneling spectra measurements.** In addition to the atomic-scale corrugations, we observed clear signature of superconductivity in tunneling spectra measurements at 4.2 K. The wide- and narrow-energy-range spectra were obtained at the area far from defects (see cross in Fig. 3a). To evaluate the superconducting gap $\Delta$ quantitatively, we analyzed tunneling spectra (Fig. 3b-d) using the Dynes formula[41] convoluted with an energy derivative of the Fermi–Dirac function $F$,

$$\frac{dI}{dV}(eV_S, T) = \int_{+\infty}^{+\infty} \left\{ \frac{dF(E + eV_S, T)}{dE} \right\} \times (a + bE) \times \text{Re} \left\{ \frac{E - i\Gamma}{\sqrt{(E - i\Gamma)^2 - \Delta^2}} \right\} dE .$$

Here, we assume that the normal-state DOS near $E_F$ has a linear dependence with energy $E$, expressed as $(a+bE)$, and $\Gamma$ is the spectral broadening factor. The fitting parameters a, b, $\Delta$, and $\Gamma$ are determined with fixed temperature $T$=4.2 K. We fitted our spectra (Fig. 3d) with a single-component isotropic pairing gap, and we obtained $\Delta$ = 1.716 ± 0.004 meV and $\Gamma$ = 0.321 ± 0.004 meV. According to the BCS gap function, assuming the same $T_c$ for surfaces and bulk (13 K), the obtained $\Delta$ value can be regarded as a gap value in the $T = 0$ K limit. Importantly, with using bulk $T_c$, the obtained $2\Delta/k_BT_c$ value of 3.0 is unexpectedly small. In contrast, the $2\Delta/k_BT_c$ values reported in polycrystalline $LiTi_2O_4$ samples range from 3.5 to 4.0 in point-contact spectroscopy[33,34] and Andreev reflection[36]. A recent report using epitaxial $LiTi_2O_4$ thin films claims $2\Delta/k_BT_c$ = 4.07 from point-contact spectroscopy[22], and thus the obtained value of 3.0 is much smaller than that of all the previous reports. Furthermore, the present value is even smaller than that for the weak coupling limit for s-wave BCS superconductivity of 3.52. We discuss later the possible origins of this unexpectedly small $2\Delta/k_BT_c$.

**Coherence length on the surface.** To further study the origin of the modified superconductivity on the surface, we investigated the value of $\xi$ from the electronic structures around a magnetic vortex core. We first analyzed the $V_s$ dependent conductance (d$I$/d$V$) map around a single vortex core by applying an external magnetic field of 1.5 T perpendicular to the surface at 4.2 K (Fig. 4a–e). At $V_s = -8$ mV and $+8$ mV, we observed uniform conductance over the scanned region (Fig. 4a and e), whereas conductance values were depressed around the center of images at $V_s = -4$ mV and $+4$ mV (Fig. 4b and d). This depressed conductance is a consequence of suppressed coherence peaks. In contrast, the conductance map at $V_s = 0$ mV clearly represents enhanced conductance in the center region (Fig. 4c). This enhanced zero bias conductance around the center region is because of pair breaking. These energy evolutions of conductance map indicate signatures of a vortex core (Fig. 4a–e), and indeed, the evolution of tunneling spectra along line A–B in Fig. 4a clearly shows a typical spatial evolution of spectral shape across a vortex core (Fig. 4f).

To evaluate $\xi$, we analyzed the zero-bias conductance data as a function of the radial distance from the vortex core center $r$ (Supplementary Fig. 3). We first extracted the zero-bias conductance $Z$ as a function of mean distance $r$. Then, we fitted the $Z(r)$ by the exponential decay function $Z(r) = Z(\infty) + A\exp(-r/\xi)$, where $A$ is a constant and $Z(\infty)$ is the normalized zero bias conductance away from the vortex core[42]. From the fitting, we obtained $\xi = 12.4 \pm 1.4$ nm (Fig. 4h), $A = 0.505 \pm 0.02$, and $Z(\infty) = 0.512 \pm 0.03$. This value is much larger than the $\xi_{GL}$ values obtained from the upper critical field ($H_{c2}$) using the formula $H_{c2} = \Phi_0/2\pi\xi_{GL}^2$ ($\Phi_0$ is the magnetic flux quanta) based on Ginzburg–Landau theory. For epitaxial LiTi$_2$O$_4$ thin films, the $\xi_{GL}$ values estimated from macroscopic measurements are 4.1–4.7 nm[22,37]. Whereas excellent agreement between $\xi$ and $\xi_{GL}$ values have been reported in other superconducting systems such as Fe-based superconductors[42,43], the values of $\xi$ obtained on the surface of LiTi$_2$O$_4$ are much larger than the estimated $\xi_{GL}$ obtained from transport measurement techniques. The deviation of $\xi$ from $\xi_{GL}$ also implies that there is a difference in superconductivity between surface and bulk.

## Discussions

To understand the superconductivity on the LiTi$_2$O$_4$ (111) surface, we performed first-principles calculations. We first calculated the bulk electronic structures to understand the triangular lattice observed in the STM images; Fig. 5a shows the calculated partial density of states for bulk. The Ti 3d states predominantly contribute near $E_F$, and the influence of Li atoms should be negligible to the STM images. Thus, the protrusions of the triangular lattice observed in the topographic image (Fig. 2) correspond to Ti atoms on the surface.

To further understand the LiTi$_2$O$_4$ (111) surface, we calculated the electronic structures of four possible bulk-cut surfaces: two-types of O-terminated, Kagome-lattice Ti-terminated, and TiLi$_2$-terminated surfaces (see four dotted lines in Fig. 5b). These surfaces were optimized structurally and the electronic states of the reconstructed surfaces were investigated. Both the two O-terminated surfaces resulted in an insulating band structure (Supplementary Figs. 4a and b), which is inconsistent with the experimental metallic tunneling spectra (Fig. 3). For the Kagome-lattice Ti-terminated surface, the simulated charge density plot (Supplementary Fig. 4c) also shows inconsistency with the experimental topographic image (Fig. 2). Consequently, neither the O-terminated nor the Kagome-lattice Ti-terminated surface reproduced the experimental results (Supplementary Fig. 4). On the other hand, TiLi$_2$-terminated surface shows metallic states, and the arrangement of the protrusions and their nearest neighbor distance (0.6 nm) observed in the STM image (Fig. 2) can be explained by the framework of Ti-triangular lattice of TiLi$_2$ layer.

We focus on the TiLi$_2$-terminated surface, and further investigate the effect of Li-layer deficiency near surface. The Li-layer deficient surfaces were examined, because Li may be deficient during depositions due to its high volatility. The TiLi$_2$-terminated surface contains a triangular lattice of Ti atoms, and two layers of Li atoms: Li atoms displaced toward the vacuum (hereafter called higher Li layer, dark green circle in Fig. 5b) and those displaced toward the bulk (hereafter called lower Li layer, light green circle in Fig. 5b). Three possible models of surface terminations exist: a stoichiometric TiLi$_2$-terminated surface, a surface without the higher Li layer (TiLi$_1$-terminated surface), and a surface without both higher and lower Li layer (TiLi$_0$-terminated surface) (Fig. 5b).

Based on the following discussion, we could exclude TiLi$_0$-terminated surface by showing that Li atoms reside at the vicinity of the topmost Ti-triangular lattice. Figure 5c shows a close-up image of the dark spots observed in the wide-area STM image (Fig. 3a). Three oval protrusions are observed, and this image indicates that a defect center is around the middle of the three oval protrusions. Considering that the ovals are at the Ti sites, and taking into account the crystal structure of spinel system, the defects could be identified as a point-Li defect. Indeed, the dark contrast around this point-Li defect, observed at a negative sample-bias voltage of -900 meV, is consistent with hole-doping nature of Li vacancy (Fig. 2a). These results demonstrate that Li atoms reside at the vicinity of the topmost Ti-triangular lattice. Accordingly, the results exclude $TiLi_0$-termination, and the surface of the films is terminated with either $TiLi_2$ or $TiLi_1$ structures. We note that the density of point-Li defects on the surface is less than 2 % of Li atoms. Thus, we speculate that the impact of the presence of point-Li defects on superconductivity can be negligible[25], unless the defects on the surfaces induce local magnetic moments.

We now compare the DOS at $E_F$, $N(E_F)$, of a bulk and that of the $TiLi_2$- and $TiLi_1$- terminated surfaces, and reveal that both surfaces have smaller $N(E_F)$ than that of the bulk. Our DFT calculations for bulk show a peak structure at $E_F$ (Fig. 5a), which is consistent with previous calculations using the linear muffin-tin orbital method[44] and full-potential linearized augmented-plane-wave[45]. The simulated peak structure at $E_F$ is also consistent with an experimental report of large normal state electronic specific heat, which is a measure of $N(E_F)$ for bulk[35]. On the $TiLi_2$- (Fig. 5d) and $TiLi_1$- (Fig. 5e) terminated surfaces, a broken lattice symmetry normal to the surface lifts the degeneracy of the $t_{2g}$ orbitals and modifies the orbital states on the surface. Compared to the bulk, we observed reduction of the $N(E_F)$ at the topmost Ti atoms for both $TiLi_2$ and $TiLi_1$ (Fig. 5f). Because the smaller $N(E_F)$ leads to lower $T_c$ according to the BCS theory, the calculation naively suggest suppressed superconductivity on the surface.

Based on the above discussions, we present coherent interpretation of the experimental results. As we indeed observed a pseudogap state experimentally (shaded red in Fig. 3 **a** and **b**), the suppressed superconductivity on the surface is a reasonable hypothesis. Furthermore, the $2\Delta/k_BT_c$ value, assuming $T_c$ to be 13 K, is determined to be 3.0, as mentioned earlier. This much small value below 3.52 is not understandable from BCS theory. A possible explanation to account for the superconductivity within the framework of BCS superconductor is to consider the presence of a non-superconducting or reduced-$T_c$ surface layer. Moreover, the large $\xi$ value can possibly be understood as the reduced $N(E_F)$. The Fermi velocity $v_F$ is proportional to $1/N(E_F)$; therefore, the reduced number $N(E_F)$ leads to an increase in $v_F$ on the surface. Thus, the large $v_F$ value on a surface increases $\xi$ based on the formula $\xi = v_F/\Delta$ derived from BCS theory. Beyond the framework of less-interacting picture, it is an interesting future subject to investigate relations between modified superconductivity, pseudogap formation, and frustration effects on this (111) surface.

In summary, we have investigated superconductivity in the atomically well-defined surface of LiTi$_2$O$_4$(111) thin films, using STM/STS and first-principles calculations. We provided spectroscopic evidence of modified superconductivity on the surface, originating from the formation of a pseudogap in the DOS. Our study has made an essential first step towards exploring superconducting phenomena emerging from spontaneous inherent atomic and electronic reconstruction on surfaces, including different film orientation[46], of superconducting cubic transition-metal oxides.

## Methods

**Samples and characterizations.** We used a low-temperature scanning tunneling microscope connected with a pulsed laser deposition (PLD) chamber. This system enables us to investigate thin-film surfaces immediately after their deposition, without exposing surfaces to air. The base pressure of the PLD chamber was $5\times10^{-11}$ Torr. Thin films of LiTi$_2$O$_4$ are grown on Nb(0.05 wt %)-doped SrTiO$_3$ (111) substrate using PLD with a KrF excimer laser (wavelength $\lambda$ =248 nm). We used Li$_4$Ti$_5$O$_{12}$ target for PLD film growth to compensate the Li loss during depositions. Substrates were annealed at 1000°C for an hour in oxygen under a partial pressure $P_{O2}$ = 5×10$^{-7}$ Torr before film depositions. Substrates were resistively heated, and their temperatures were monitored using a pyrometer. We deposited film at a substrate temperature $T_s$ = 300°C under $P_{O2}$ = 5×10$^{-7}$ Torr followed by post-deposition annealing at 600°C in UHV for one hour. During thin film depositions, pulse repetition and fluence were set at 2 Hz and 1 J/cm$^2$, respectively. Film thickness, measured *ex situ* using a DEKTAK 3030ST mechanical profiler, was determined to be about 100 nm. All STM experiments were performed at liquid-He temperature (4.2 K). We obtained differential conductance values (d$I$/d$V$) from the numerical derivatives of $I$ (tunneling current) - $V$ (voltage) curves. For ex situ characterization, we measured X-ray diffraction pattern and temperature dependence of magnetization after in situ STM/STS measurements.

**First principles calculations.** For the first-principles calculations, we used DFT with the code Quantum ESPRESSO, with the generalized gradient approximation and ultra-soft pseudopotential scheme[47–51]. Cutoff energies for the Kohn–Sham orbitals and charge density of 32.5 Ry and 45 Ry are imposed. The Brillouin-zone summation is evaluated using 3 × 3 × 1 and 12 × 12 × 1 k-point sampling for the structure optimizations and the density of states calculations, respectively. For the density of states broadening, we applied a simple Gaussian broadening method with broadening parameter of 0.005 Ry. Convergence criteria of the structure optimization are 10×10$^{-3}$ for forces and 10×10$^{-4}$ for energy. For the calculation of the near surface atomic/electronic structures, we use the symmetric slab model with 4 unit cells along the (111) direction on each side. In the structure optimization, the topmost surface structures are initiated from bulk cuts. The atomic positions of the Ti atoms at the center of the symmetric slab model along the (111) direction are fixed during optimization. Spin degrees of freedom and electron–electron correlations are not included in the calculation.

**Data availability**

The data that support the findings of this study are available from the corresponding author upon reasonable request.

**Acknowledgement**

We thank Y. Takagi and K. Yamamoto for experimental assistance, and T. Hanaguri, T. Machida, Y. Yoshida, H. Kawasoko, and K. Sato for useful discussions. Y. O. acknowledges funding from JSPS KAKENHI Grant Nos. 26707016 and 25886004. T. H. acknowledges funding from JSPS KAKENHI Grant Nos. 26246022, 26106502, 26108702, 26610092, JST-PRESTO, and JST-CREST program. This work was supported by the World Premier Research Center Initiative (WPI), promoted by the Ministry of Education, Culture, Sports, Science and Technology (MEXT) of Japan. We thank P. Han for the critical reading of this manuscript.

**Author Contributions**

Y. O. pursued film growth, STM/STS measurements, and data analysis. Y. A performed the first-principles calculations. Y. O., R. S., S. S., and T. H. discussed the experimental results. Y. A., E. M., and S. W. discussed the theoretical calculations. All authors discussed the conclusions of this paper. Y. O. and T. H. wrote the paper.

**Competing financial interests**

The authors declare no competing financial interests.

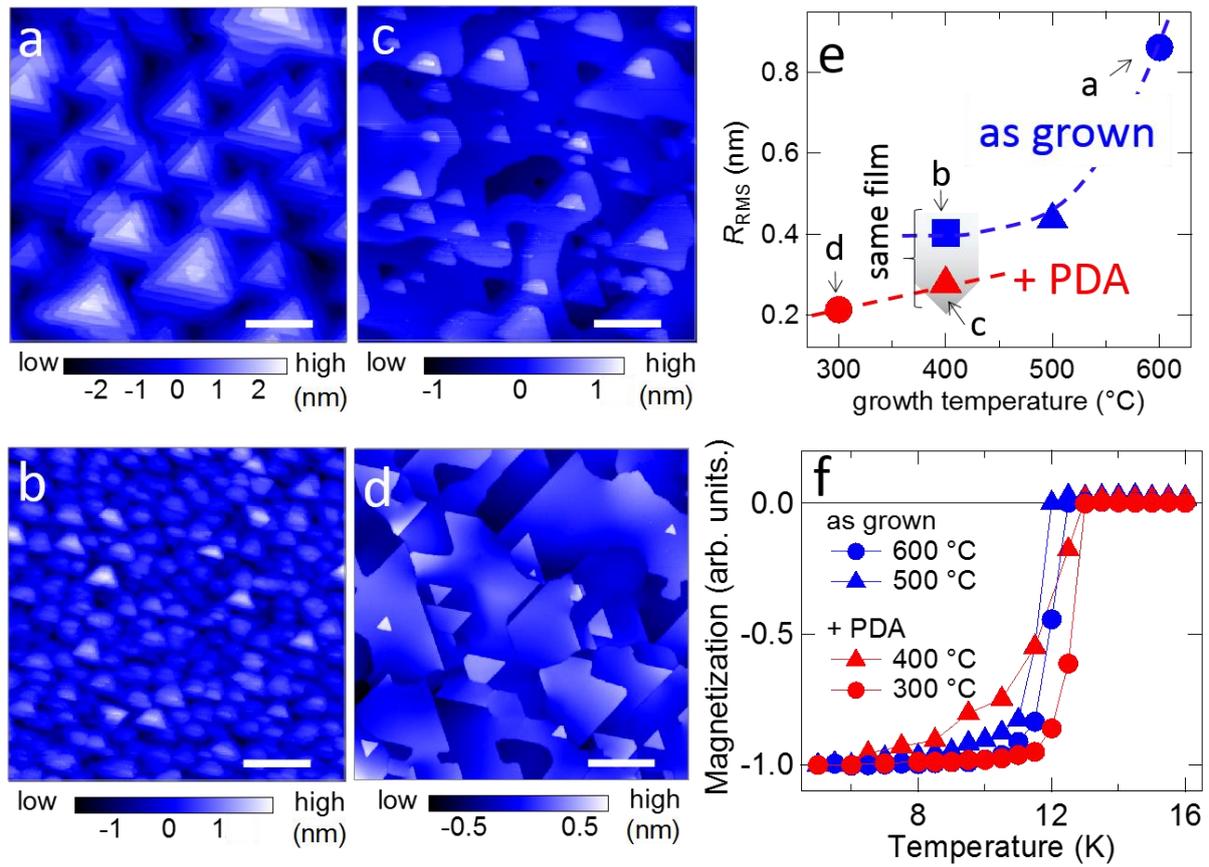

**Figure 1 | Surface topographies and superconducting critical temperatures.** (**a, b**) Scanning tunneling microscope (STM) topographic images of as-deposited thin film at substrate temperature of 600 °C (**a**) and 400 °C (**b**). (**c, d**) STM images after post-deposition annealing (PDA) for films deposited at 400 °C (**c**), and 300 °C (**d**). Note that **b** and **c** are taken with using the same film. Figs. **a**-**c** are obtained at 77 K and **d** is obtained at 4.2 K (All the STM images were observed at a sample-bias voltage of + 300 mV and a tunneling current is about 10 pA). The scale bars in (**a**-**d**) indicate 80 nm. (**e**) Growth temperature dependence of root mean square of surface roughness ($R_{RMS}$) values: as grown samples (blue symbols), and after PDA (red symbols). The value of $R_{RMS}$ is evaluated from topographic images observed at a sample-bias voltage of + 300 mV and a tunneling current of 10 pA (scan area of 400 nm). (**f**) Temperature dependence of the field-cooled dc magnetic susceptibility for the $LiTi_2O_4$ films in $H$ = 50 Oe, which was applied parallel to the (111) plane. Clear diamagnetism is observed.

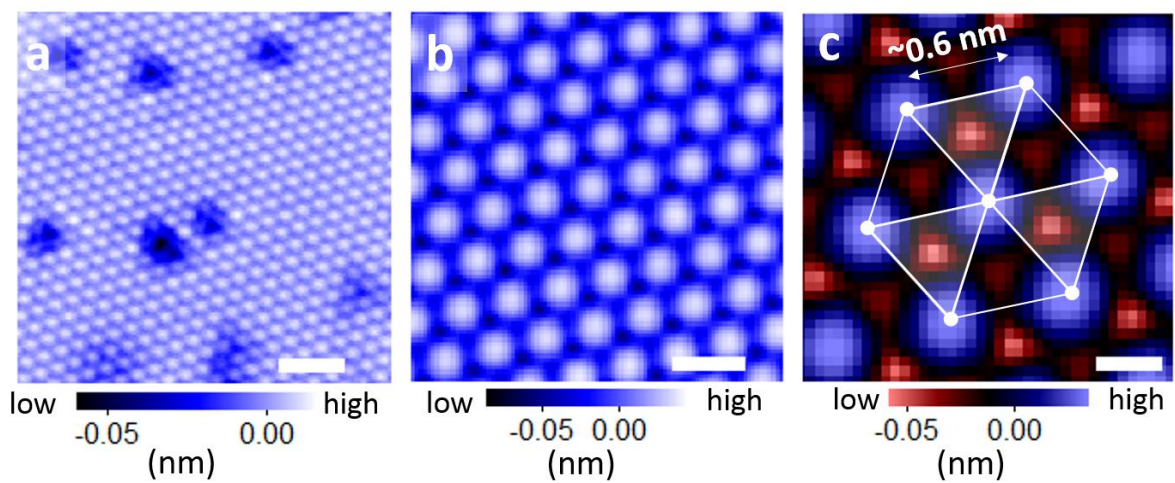

**Figure 2 | Typical topographic images on a terrace.** (**a**) Empty-state scanning tunneling microscope (STM) image of LiTi$_2$O$_4$ (111) surface (11.6 nm × 11.6 nm, sample bias voltage $V_s$ of -900 mV, tunneling current $I_{set}$ of 30 pA). (**b**) Filled-state STM image (4 nm ×4 nm, $V_s$ =+ 30 mV, $I_{set}$ = 30 pA). (**c**) Zoomed-up image (1.7 nm ×1.7 nm, $V_s$ = +30 mV, $I_{set}$ = 30 pA) of b. The image shows three-fold symmetry representing the spinel crystal structure. The scale bars in **a**, **b**, and **c** indicate 2 nm, 0.8 nm, and 0.3 nm, respectively.

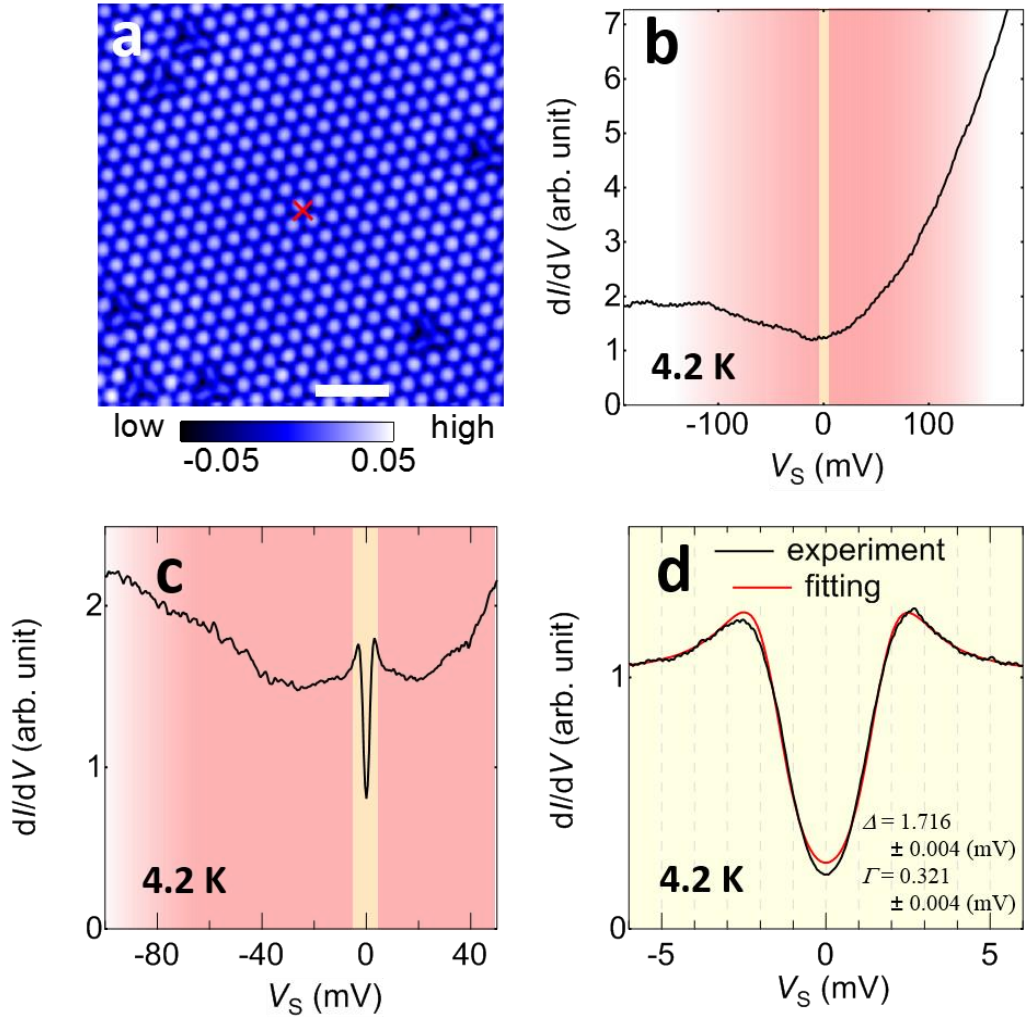

**Figure 3 | Tunneling spectra obtained away from defects.** (**a**) A topographic image obtained with a sample bias voltage $V_s$ of + 30 mV and a tunneling current of 30 pA. The red cross in **a** indicates the position where the spectra shown in **b**, **c**, and **d**. were obtained. (**b**) (**c**) Tunneling spectrum (d$I$/d$V$) obtained at a wide energy region ($V_s$ of −190 mV to +190 mV and −100 mV to +50 mV, for **a** and **b**, respectively.) (**c**) High-resolution tunneling spectrum for $V_s$ between ± 6 mV near the Fermi energy. Experimental curve (black dots) and fitted curve (red line) are shown in the same figure. See main body for the details of the fitting procedure. The energy windows of numerical derivative to obtain conductance spectrum were 15 mV, 1.5 mV and 0.3 mV for **a, b** and **c**, respectively. The yellow region represents the $V_s$ range of -6 mV to 6 mV. All the images and spectra were obtained at 4.2 K.

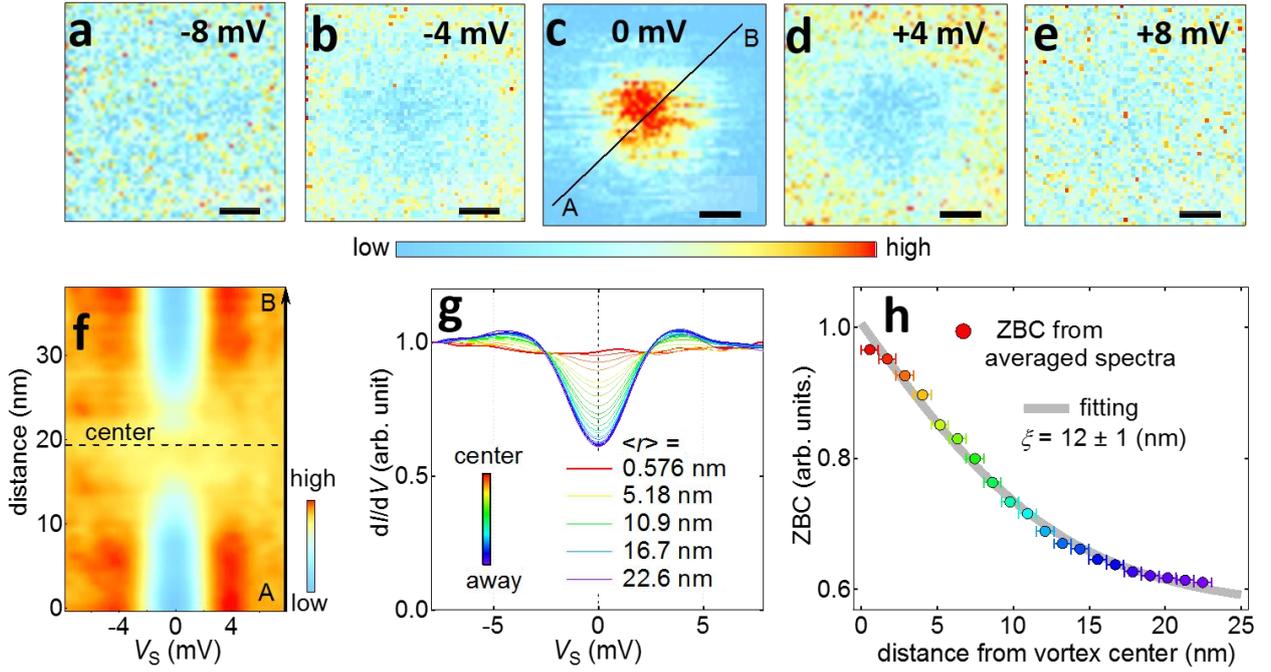

**Figure 4 | Spectral evolution around a vortex core.** A single vortex core is investigated at the temperature of 4.2 K with applying a magnetic field of 1.5 T normal to the surface. (**a–e**) Conductance mappings around an isolated single vortex core with various sample bias voltages. An scanning tunneling microscope tip is stabilized at a tunneling current of 30 pA and a sample bias voltage of −10 mV. The energy window of numerical derivative to obtain conductance spectrum is 1 mV. The scale bars in (**a-e**) indicate 6 nm. (**f**) Spatial evolution of tunneling spectra across the vortex center (the line is shown in **c**). (**g**) Averaged spectra as function of distance from vortex center $<r>$. (**h**) Zero bias conductance (ZBC) obtained from **g**. Coherence length of ~12 nm is obtained by fitting the ZBC (see main text for details). 3600 spectra (60 × 60) were taken with equal spacing in the region (**a-e**). We classified the area of Fig. 4**a–e** into 20 regions based on the distance from vortex center (see also Supplementary Fig. 3). Interval of $<r>$, which is 5.758 nm, is set as error bars for x-axis in **h**.

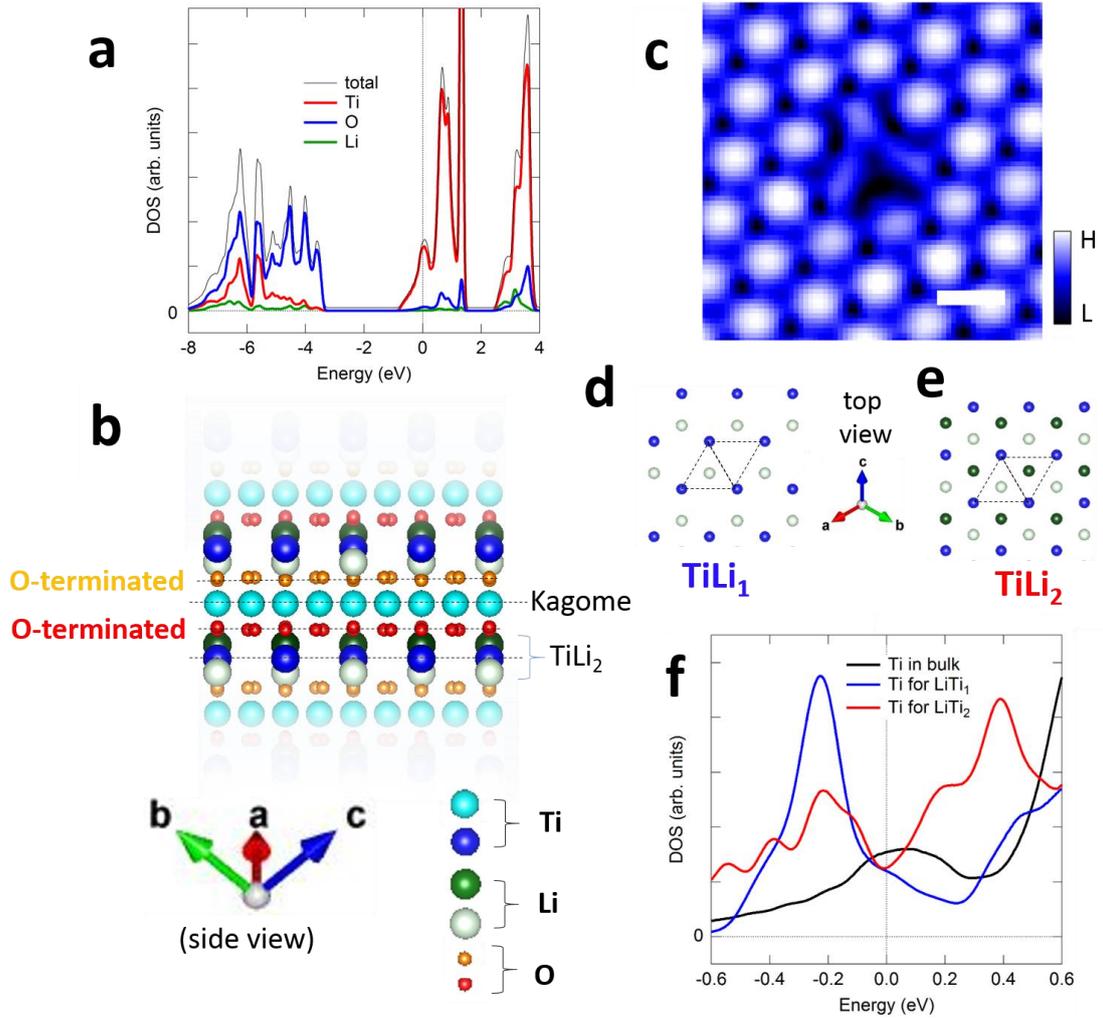

**Figure 5 | Comparison between bulk and surface electronic states based on density functional theory (DFT) calculations.** (**a**) Calculated density of states (DOS) for bulk $LiTi_2O_4$. (**b**) Crystal structure with bulk continuum, together with four bulk-cut planes represented by broken lines. Vertical axis is along the (111) crystal orientation. (**c**) The topographic image of a defect on the surface (a sample-bias voltage of +30 mV, a tunneling current of 30 pA). The scale bar in (**c**) indicates 0.6 nm. Top view of $TiL_2$- (**d**) and $TiLi_1$- (**e**) terminated surfaces. See (**b**) for the color of the circles. (**f**) Calculated DOS for the top-most Ti atoms in $TiL_2$- and $TiLi_1$-terminated surfaces. DOS for bulk is shown again for clarity. Here, 0 on the horizontal axes in **a** and **f** correspond to the Fermi energy.

23